\documentclass[prb,twocolumn,superscriptaddress]{revtex4-2}
\usepackage[dvipdfmx]{graphicx} 
\usepackage{dcolumn}
\usepackage{bm}
\usepackage{color}
\usepackage{multirow}
\usepackage{lineno}

\graphicspath{{fig/}}

\begin{document}

\preprint{APS/123-QED}

\title{
Theoretical analysis on the possibility of superconductivity in a trilayer Ruddlesden-Popper nickelate La$_4$Ni$_3$O$_{10}$ under pressure and its experimental examination : comparison with La$_3$Ni$_2$O$_7$
}

\author{Hirofumi Sakakibara}
\thanks{These five authors contributed equally}

\affiliation{Advanced Mechanical and Electronic System Research Center(AMES), Faculty of Engineering, Tottori University, 4-10 Koyama-cho, Tottori, Tottori 680-8552, Japan}

\author{Masayuki Ochi}
\thanks{These five authors contributed equally}

\affiliation{Department of Physics, Osaka University, 1-1 Machikaneyama-cho, Toyonaka, Osaka 560-0043, Japan}
\affiliation{Forefront Research Center, Osaka University, 1-1 Machikaneyama-cho, Toyonaka, Osaka 560-0043, Japan}

\author{Hibiki Nagata}
\thanks{These five authors contributed equally}

\affiliation{MANA, National Institute for Materials Science (NIMS), 1-2-1 Sengen, Tsukuba 305-0047 Japan}
  \affiliation{University of Tsukuba, 1-1-1 Tennodai, Tsukuba, Ibaraki 305-8577, Japan}

\author{Yuta Ueki}
\thanks{These five authors contributed equally}
\affiliation{MANA, National Institute for Materials Science (NIMS), 1-2-1 Sengen, Tsukuba 305-0047 Japan}
  \affiliation{University of Tsukuba, 1-1-1 Tennodai, Tsukuba, Ibaraki 305-8577, Japan}

\author{Hiroya Sakurai}
\thanks{These five authors contributed equally}

\affiliation{MANA, National Institute for Materials Science (NIMS), 1-2-1 Sengen, Tsukuba 305-0047 Japan}

\author{Ryo Matsumoto}
\affiliation{MANA, National Institute for Materials Science (NIMS), 1-2-1 Sengen, Tsukuba 305-0047 Japan}

\author{Kensei Terashima}
\affiliation{MANA, National Institute for Materials Science (NIMS), 1-2-1 Sengen, Tsukuba 305-0047 Japan}

\author{Keisuke Hirose}
\affiliation{Department of Molecular Chemistry and Biochemistry, Doshisha University, 1-3 Tataramiyakotani, Kyo-Tanabe 610-0321, Japan}
\author{Hiroto Ohta} 
\affiliation{Department of Molecular Chemistry and Biochemistry, Doshisha University, 1-3 Tataramiyakotani, Kyo-Tanabe 610-0321, Japan}
\author{Masaki Kato}
\affiliation{Department of Molecular Chemistry and Biochemistry, Doshisha University, 1-3 Tataramiyakotani, Kyo-Tanabe 610-0321, Japan}

\author{Yoshihiko Takano}
\email{TAKANO.Yoshihiko@nims.go.jp}
\affiliation{MANA, National Institute for Materials Science (NIMS), 1-2-1 Sengen, Tsukuba 305-0047 Japan}
\affiliation{University of Tsukuba, 1-1-1 Tennodai, Tsukuba, Ibaraki 305-8577, Japan}
\author{Kazuhiko Kuroki}
\email{kuroki@presto.phys.sci.osaka-u.ac.jp}
\affiliation{Department of Physics, Osaka University, 1-1 Machikaneyama-cho, Toyonaka, Osaka 560-0043, Japan}

\date{\today}

\begin{abstract}
We study the possibility of superconductivity in a trilayer Ruddlesden-Popper nickelate La$_4$Ni$_3$O$_{10}$ under pressure both theoretically and experimentally, making comparison with the recently discovered high $T_c$ superconductor La$_3$Ni$_2$O$_7$, a bilayer nickelate. Through DFT calculations, we find that a structural phase transition from monoclinic to tetragonal takes place around 10 -- 15  GPa. Using the tetragonal crystal structure, we theoretically investigate the possibility of superconductivity, where a combination of fluctuation exchange approximation and linearized Eliashberg equation is applied to a six-orbital model constructed from first principles band calculation. The obtained results suggests that La$_4$Ni$_3$O$_{10}$ may also become superconducting under high pressure with $T_c$ comparable to some cuprates, although it is not as high as La$_3$Ni$_2$O$_7$. 
We also perform experimental studies using our polycrystalline samples of La$_3$Ni$_2$O$_{7.01}$ and La$_4$Ni$_3$O$_{9.99}$. The superconducting transition of La$_3$Ni$_2$O$_{7.01}$, with a maximum onset $T_c$ of  67.0 K  at a pressure of  26.5  GPa, is confirmed by a drop in the electrical resistance, as well as the magnetic field dependence of the resistance. Quite interestingly, similar temperature and magnetic field dependencies of the resistance are observed also for La$_4$Ni$_3$O$_{9.99}$, where a drop in the resistance is observed at lower temperatures compared to La$_3$Ni$_2$O$_{7.01}$, under pressures of 32.8 GPa and above. Given the theoretical expectation, the reduction in the resistance can most likely be attributed to the occurrence of superconductivity in La$_4$Ni$_3$O$_{9.99}$. The temperature at which the resistance deviates from a linear behavior, considered as the onset $T_c$, monotonically increases up to 23 K at 79.2 GPa, which is opposite to the pressure dependence of $T_c$ in La$_3$Ni$_2$O$_{7.01}$.  
\end{abstract}

\maketitle

\section{Introduction}

Even vs. odd number effects have often been issues of interest in various fields of physics. A famous example in condensed matter physics is the Haldane's conjecture, which states that in a spin-$N/2$ antiferromagnetic Heisenberg chain, the excitation spectrum is gapless and the spin-spin correlation exhibits a slow (algebraic) decay when $N$ is an odd number, whereas when $N$ is even, there is a gap in the excitation and the spin correlation decays exponentially~\cite{Haldane}.  Schulz discussed the relation between spin-$N/2$ antiferromagnetic Heisenberg chain and $N$ coupled spin-1/2 chains~\cite{Schulz,SchulzRev}, namely, $N$-leg ladders, and argued that when $N$ is even, the system is gapful, while the system becomes gapless when $N$ is odd.  In the 1990's, following the studies by  Dagotto {\it et al.}\cite{DagottoScalapino} and by Rice {\it et al.}\cite{Rice}, spin-1/2 antiferromagnetic $N$-leg ladders were intensively studied both theoretically and experimentally~\cite{DagottoRice}. Due to the opening of the spin gap in the case of $N=$ even, the possibility of superconductivity in doped even-leg (two-leg in particular) Hubbard and $t$-$J$ models was widely investigated both analytically and numerically~\cite{DagottoScalapino,Rice,DagottoRice}. Indeed, a cuprate that contains two-leg ladders was found to superconduct under pressure~\cite{Akimitsu}.  

Here, an interesting problem arises regarding the possibility of superconductivity in doped {\it odd}-number-leg ladders. One might expect absence of superconductivity due to the absence of the spin gap, but one of the present authors and his colleagues showed that superconductivity can still take place in the three-leg Hubbard ladder~\cite{Kimura}, based on the weak coupling analysis that shows two out of three spin modes are gapped~\cite{Arrigoni,Arrigoni2}.  A similar result was also obtained in Ref.~\cite{SchulzRev}, and the weak coupling analysis was soon extended to $N$-leg ladders~\cite{LinBalentsFisher}.

A two dimensional analogue of the two-leg ladder model, namely, the bilayer model, typically on a square lattice, has also been a target  of theoretical interest from the perspective of unconventional superconductivity. When two square lattices are coupled by a large vertical hopping $t_\perp$ or a strong vertical magnetic coupling $J_\perp$,  opening of a spin gap is expected, and various studies have shown strong enhancement of interlayer pairing superconductivity~\cite{DagottoScalapino,Bulut,KA,Maier,MaierScalapino,Nakata,MaierScalapino2,KarakuzuMaier,Matsumoto2,DKato}. Motivated by the very strong enhancement of superconductivity near half filling found in some studies~\cite{KA,Maier},  one of the present authors investigated the possibility of realizing the bilayer Hubbard model with large $t_\perp$ in actual materials, and ended up with a bilayer Ruddlesden-Popper nickelate La$_3$Ni$_2$O$_7$~\cite{Nakata}. The key factor there was that the Ni $3d_{3z^2-r^2}$ orbitals, which are elongated in the out-of-plane direction and hence enhance $t_\perp$, are close to half filling in this material. Quite recently, superconductivity in the very La$_3$Ni$_2$O$_7$ with a maximum $T_c$ of 80 K has been discovered under high pressure~\cite{MWang}, which has sparked a vast wave of interest both experimentally~\cite{2307.02950,2307.09865,2307.14819,2309.01148} and theoretically~\cite{2306.07837,2306.03706,WernercRPA,2307.05662,2307.16697,2307.16873,2307.15706,LuoPRL,2307.10144,2307.06806,2307.14965,2307.07154,2306.03231,2307.15276,2306.05121, 2308.01176, 2308.06771, 2306.07275,2309.06173,2309.05726,2308.16564,2308.11614,2308.11195,2308.09698,2308.09044,2308.07651,2308.07386,2309.13040,2309.15078} including ours~\cite{sakakibarala327}. 

Given the recent developments in the bilayer La$_3$Ni$_2$O$_7$,  and also the previous studies regarding even vs. odd-leg ladders as described above,  it is natural to consider investigating the possibility of superconductivity in a trilayer Ruddlesden-Popper nickelate La$_4$Ni$_3$O$_{10}$. Since La$_4$Ni$_3$O$_{10}$ is not superconducting at ambient pressure, we will consider its possibility under pressure  as in La$_3$Ni$_2$O$_7$ both theoretically and experimentally. Since superconductivity in La$_3$Ni$_2$O$_7$ appears to occur when the symmetry of the crystal structure becomes (close to) tetragonal under high pressure~\cite{MWang}, we first investigate theoretically whether the crystallographic symmetry of La$_4$Ni$_3$O$_{10}$ becomes tetragonal under high pressure. Indeed, we find that a structural phase transition from monoclinic to tetragonal takes place around 10 -- 15  GPa. Adopting a hypothesis that superconductivity in La$_4$Ni$_3$O$_{10}$, if any, occurs when the crystallographic symmetry is tetragonal, we investigate the possibility of superconductivity using the crystal structure obtained at 40 GPa, which is safely in the tetragonal phase regime. A combination of fluctuation exchange approximation (FLEX) and linearized Eliashberg equation is applied to a six-orbital (2 orbitals $\times$ 3 layers) model constructed from first principles band calculation. Our calculation results suggest that La$_4$Ni$_3$O$_{10}$ may also become superconducting under high pressure with $T_c$ comparable to some cuprates, although it is not as high as La$_3$Ni$_2$O$_7$.

We also present experimental results for our polycrystalline samples of La$_3$Ni$_2$O$_{7.01}$  and La$_4$Ni$_3$O$_{9.99}$. The superconducting transition of La$_3$Ni$_2$O$_{7.01}$, with a maximum onset $T_c$ of 67.0  K at a pressure of 26.5 GPa, is confirmed by a drop in the electrical resistance, as well as the magnetic field dependence of the resistance. Quite interestingly, similar temperature and magnetic field dependencies of the resistance are observed also for La$_4$Ni$_3$O$_{9.99}$, where a drop in the resistance is observed at lower temperatures compared to La$_3$Ni$_2$O$_{7.01}$, under pressures of 32.8 GPa and above. Given the theoretical expectation, the reduction in the resistance can most likely be attributed to the occurrence of superconductivity in La$_4$Ni$_3$O$_{9.99}$.

\section{Theoretical Method}
For density functional theory (DFT) calculation, we use the PBEsol exchange-correlation functional~\cite{PBEsol} and the projector augmented wave method~\cite{paw} as implemented in {\it Vienna ab initio Simulation Package} (VASP)~\cite{vasp1,vasp2,vasp3,vasp4}. 
Core-electron states in PAW potentials are [Kr]$4d^{10}$, [Ar], [He] for La, Ni, O, respectively.
We use a plane-wave cutoff energy of 600 eV for Kohn-Sham orbitals without including the spin-orbit coupling for simplicity.

We perform structural optimization until the Hellmann-Feynman force becomes less than 0.01 eV \AA$^{-1}$ for each atom using an $8\times 8\times 2$ ${\bm k}$-mesh.
To verify the stability of the optimized crystal structure under pressure, we calculate the phonon dispersion using the finite displacement method as implemented in the \textsc{Phonopy}~\cite{phonopy} software in combination with VASP.
We use a $3\times 3\times 1$ ${\bm q}$-mesh for a conventional tetragonal unit cell containing 34 atoms.
For a $3\times 3\times 1$ supercell used for finite-displacement calculations, we use a $4\times 4\times 2$ ${\bm k}$-mesh.

To discuss superconductivity, we extract Ni-$d_{x^2-y^2}$ and $d_{3z^2-r^2}$ Wannier orbitals using \textsc{Wannier90} software~\cite{Wannier1,Wannier2,Wannier90}.
For this purpose, we use the tetragonal crystal structure under the pressure of 40 GPa obtained by our calculation. We use an $12\times 12\times 12$ ${\bm k}$-mesh for a primitive unit cell. 
We adopt FLEX~\cite{Bickers,Bickers1991} in order to take into account the effect of electron correlation as was done in Ref.~\cite{sakakibarala327} for La$_3$Ni$_2$O$_7$.
As for the interaction term of the Hamiltonian, we take the on-site interactions, namely, intraorbital(interorbital) Coulomb interactions $U$($U'$), 
Hund's coupling $J$, and pair hopping $J'$.
We assume the orbital rotational symmetry, and take the same value of $U$ for the $d_{x^2-y^2}$ and the $d_{3z^2-r^2}$ orbitals,
and $U'=U-2J, J=J'$.  As a typical value of the interactions, we take $U=3$ eV,   
$J=J'=0.3$ eV and $U'=U-2J=2.4$ eV. 
These interaction values are the same as those adopted in our study for La$_3$Ni$_2$O$_7$, which can be considered as typical values for 3$d$-transition-metal oxides~\cite{note-int}.
We calculate the self-energy induced by the spin-fluctuation formulated as shown in the literatures~\cite{Lichtenstein,mFLEX1,mFLEX2} in a self-consistent calculation.
The real part of the self-energy at the lowest Matsubara frequency is subtracted in the same manner with Ref.~\cite{Ikeda_omega0} to maintain the band structure around the Fermi level obtained by first-principles calculation.

We use the linearized Eliashberg equation to study the possibility of superconductivity,  also as in Ref.~\cite{sakakibarala327} for La$_3$Ni$_2$O$_7$. The renormalized Green's functions obtained by FLEX are plugged into this equation. Also, the pairing interaction kernel in this equation is obtained from the FLEX Green's function as a purely electronic one (i.e. phonon-mediated pairing interaction is not considered), which is mainly dominated by spin fluctuations in the present case.
Since the the eigenvalue $\lambda$ of the Eliashberg equation monotonically increases upon lowering the temperature, and reaches unity at $T=T_c$, 
we adopt $\lambda$ calculated at a fixed temperature, $T=0.01$ eV as a measure of superconductivity. For convenience, we will call the eigenfunction (with the largest eigenvalue) of the linearized Eliashberg equation at the lowest Matsubara frequency $i\omega$(=$i\pi k_{\rm B}T$) the ``superconducting gap function''. We take a 16$\times$16$\times$4 $k$-point mesh and 2048 Matsubara frequencies for the FLEX calculation.

\section{Experimental method}
Polycrystalline samples of La$_3$Ni$_2$O$_{7.01}$ and La$_4$Ni$_3$O$_{9.99}$ were employed for electrical resistance measurements under high-pressure conditions. These were synthesized from La$_2$O$_3$ and NiO, and characterized by powder X-ray diffraction and thermogravimetry to find that they were of single phases with the chemical compositions. The details of the synthesis and characterization will be described elsewhere~\cite{Sakuraiprep}. High pressure was generated with Diamond Anvil Cell (DAC) with boron-doped diamond electrodes designed for four-terminal resistance measurement ~\cite{RM2016RSI,RM2018APEX}. Cubic boron nitride powder was used as a pressure-transmitting medium. Applied pressure was estimated by fluorescence of ruby placed near the sample up to 20 GPa~\cite{Piermarni}. For pressure beyond 20 GPa, applied pressure was estimated by a Raman spectrum from a culet of the top diamond anvil obtained by Raman Microscope (Renishaw)~\cite{Akahama}. Details of the cell configuration will be  described in the literatures~\cite{RM2018STAM,RM2017JJAP}. Temperature and magnetic fields were controlled by PPMS (Quantum Design).

\section{Theoretical Results--Crystal Structure} 
\begin{figure*}
\begin{center}
\includegraphics[width=15 cm]{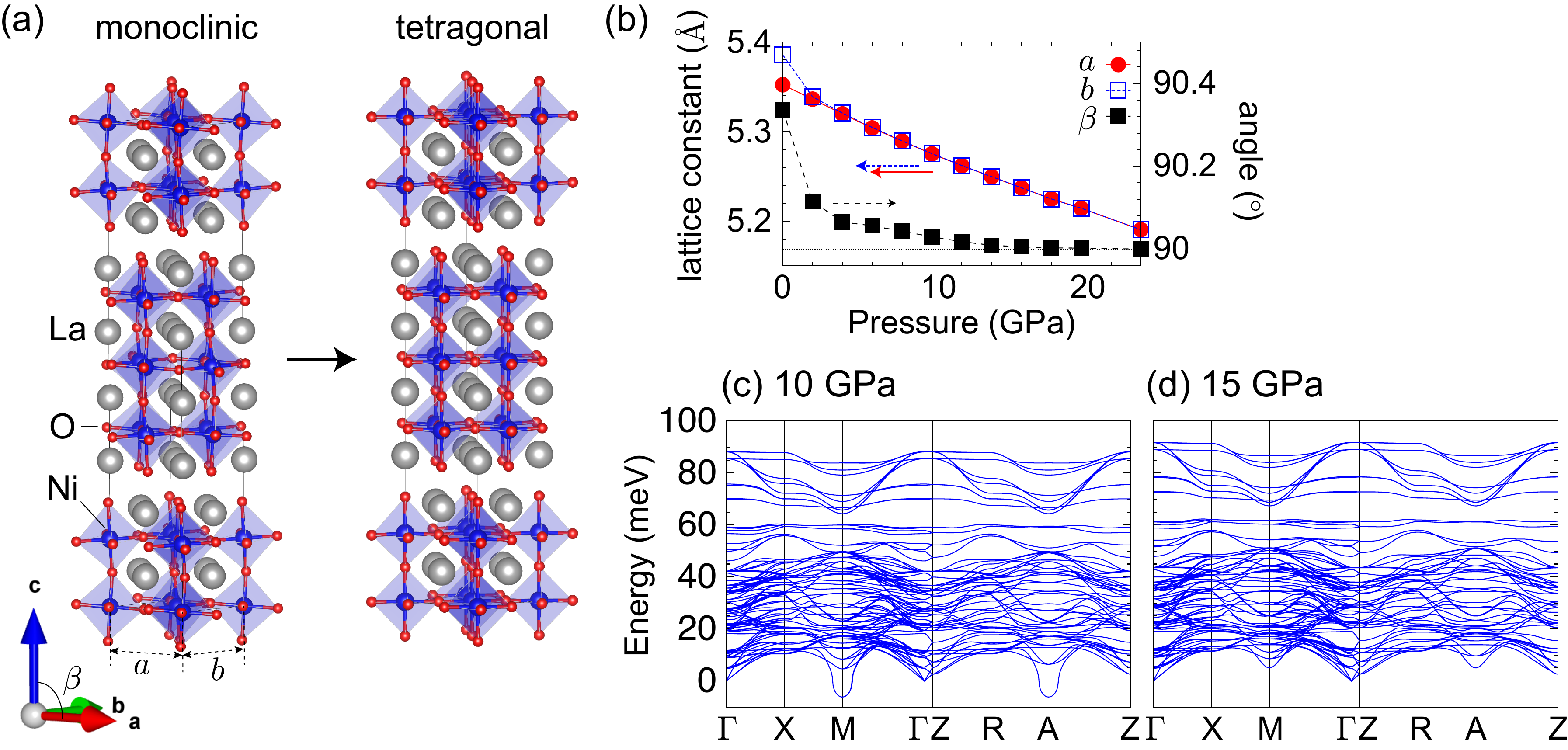}
\caption{(a) Optimized crystal structures of La$_4$Ni$_3$O$_{10}$ red in the monoclinic and tetragonal phases. Gray, blue, and red spheres denote La, Ni, and O atoms, respectively. Crystal structure was depicted using the VESTA software~\cite{VESTA}. (b) Lattice parameters $a$, $b$, and $\beta$ obtained by structural optimization with the $P2_1/a \ (Z=4)$ space group under pressure. (c)(d)Phonon dispersion of the tetragonal $I4/mmm$ structure under the pressure of 10 and 15 GPa, respectively.}
\label{fig:fp}
\end{center}
\end{figure*}
We optimized the crystal structure under pressure by DFT calculation.
Here, we note that several different space groups have been reported in literature for La$_4$Ni$_3$O$_{10}$ at ambient pressure, 
$Fmmm$~\cite{La4310_strct} (or $Imm2$ considering a symmetry lowering mentioned in this study, as pointed out in Ref.~\onlinecite{La4310_PRMater}),
$Cmce$ ($Bmab$)~\cite{La4310_strct2, La4310_strct3, La4310_PRMater},
$P2_1/a \ (Z=4)$ (i.e., $P2_1/a$ containing 4 formula units in the unit cell)~\cite{La4310_Co, La4310_mono,La4310_Aldoped} as was also reported for Nd$_4$Ni$_3$O$_{10-\delta}$~\cite{Nd4310, Nd4310_2},
and $P2_1/a \ (Z=2)$~\cite{LaPr4310, La4310_PRMater} as was also reported for Pr$_4$Ni$_3$O$_{10}$~\cite{Pr4310}.
It was pointed out that samples were mixed phases of orthorhombic and monoclinic symmetry~\cite{La4310_ARPES, La4310_PRMater}.
The stability of several phases, $Cmce$, $Pbca$, and two types of $P2_1/a$, was also investigated by DFT calculation~\cite{La4310_calc}.
Due to the complexity of the situation, we first optimized a crystal structure under pressure assuming a monoclinic space group of $P2_1/a \ (Z=4)$, which has a relatively low symmetry among these candidates and is a subgroup of some of the candidates, and next checked the stability of the obtained structure under pressure by phonon calculation.

By the structural optimization of an $P2_1/a \ (Z=4)$ structure under pressure, we found that the structural phase transition from monoclinic to tetragonal takes place as shown in Fig.~\ref{fig:fp}(a).
In fact, lattice parameters relevant to the monoclinic-tetragonal transition, $a$, $b$, and $\beta$, were shown in Fig.~\ref{fig:fp}(b), where $a=b$ and $\beta = 90^{\circ}$ were realized at around 10 -- 15 GPa. We also confirmed that the obtained tetragonal structure belongs to the $I4/mmm$ space group by checking atomic coordinates.
Then, we performed phonon calculation for the tetragonal $I4/mmm$ structure under the pressure of 10 and 15 GPa as shown in Figs.~\ref{fig:fp}(c)--(d).
While the calculated phonon dispersion has an imaginary mode at 10 GPa, this instability disappears at 15 GPa,
by which we confirmed that the structural transition takes place at around 10 -- 15 GPa and the tetragonal structure is dynamically stable above 15 GPa.
Our calculation shows that several tens of gigapascal pressure gives rise to a tetragonal structure, as in La$_3$Ni$_2$O$_7$~\cite{MWang}.

\begin{figure*}
\begin{center}
\includegraphics[width=18cm]{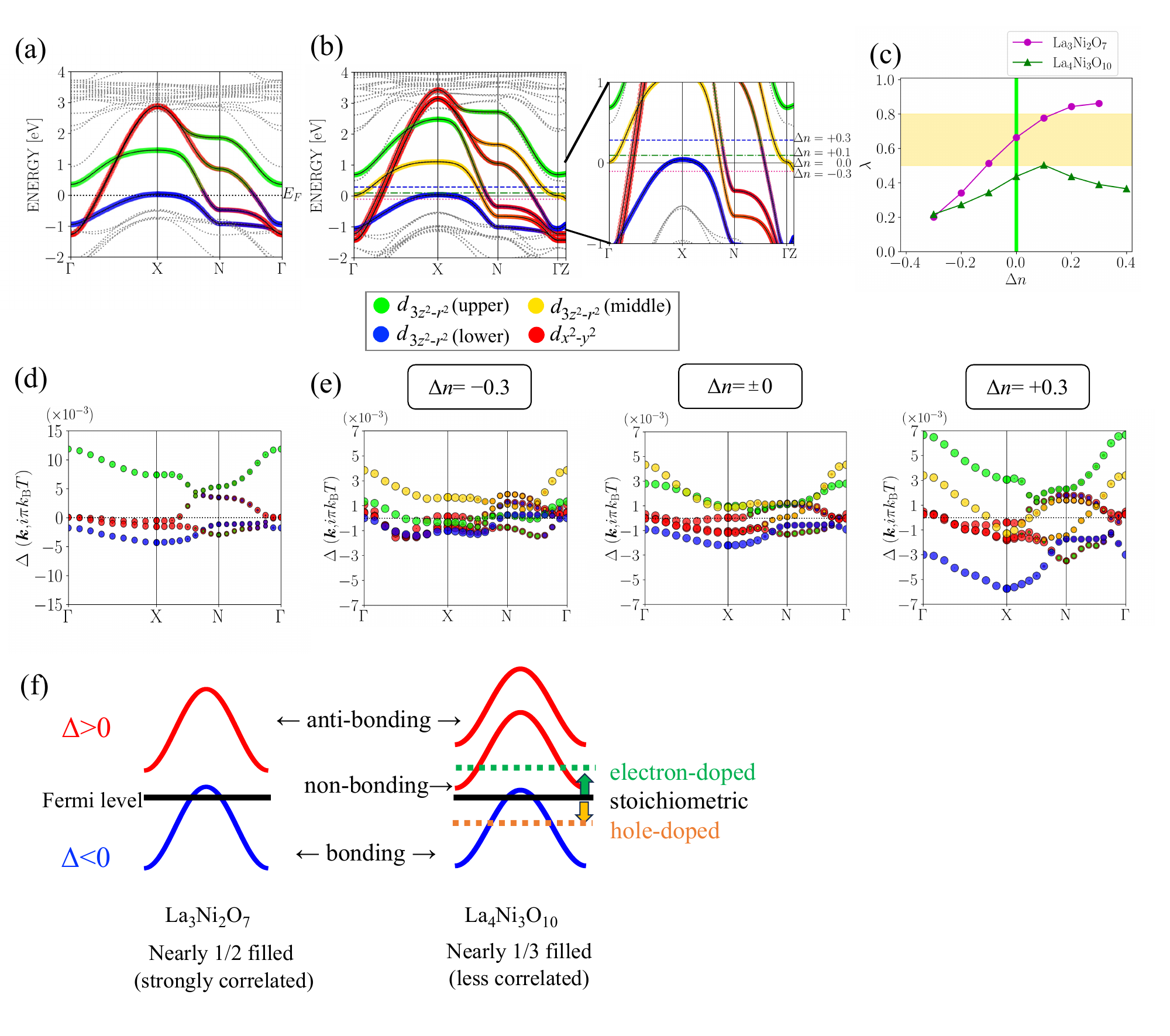}
\caption{(a) The band structure of the four-orbital model of La$_3$Ni$_2$O$_7$~\cite{sakakibarala327}. (b) The band structure of the six-orbital model of La$_4$Ni$_3$O$_{10}$. On the right is a blowup near the Fermi level, together with the Fermi level for various band fillings. (c) The eigenvalue of the Eliashberg equation for La$_4$Ni$_3$O$_{10}$ and La$_3$Ni$_2$O$_7$~\cite{sakakibarala327} plotted against the band filling measured from stoichiometry. The yellow hatched region indicates the range of $\lambda$ obtained by the same method for various cuprates~\cite{sakakibara1,sakakibara2,sakakibara3}. (d) Superconducting gap functions of  La$_3$Ni$_2$O$_7$ for $\Delta n=0$~\cite{sakakibarala327}.
  (e) Superconducting gap functions of La$_4$Ni$_3$O$_{10}$ for $\Delta n=-0.3$ (left), $\Delta n=0$ (center),  $\Delta n=+0.3$ (right).
  (f) Schematic picture of the superconducting gap functions of La$_3$Ni$_2$O$_7$ and La$_4$Ni$_3$O$_{10}$.
The strength of Wannier orbital characters are presented in panels (a)(b)(d)(e) with the thickness/radius of the color coded line/circles,
  where the ``sum'' of the $d_{x^2-y^2}$($d_{3z^2-r^2}$) orbital characters among inner- and  outer-layer is indicated by red (green, blue, and yellow). 
  The yellow color is used for indicating the $d_{3z^2-r^2}$ characters in the case that the eigenstate contains the inner-layer components less than 5\%.
 Otherwise, either green or blue is used for the $d_{3z^2-r^2}$ components, depending on whether the band energy is above or below 0.2 eV.
  In this way, the colors blue, yellow, and green represent bonding, non-bonding, and anti-bonding $d_{3z^2-r^2}$ bands, respectively.
  All the calculation results for La$_4$Ni$_3$O$_{10}$ are obtained using the crystal structure at 40 GPa.
}
\label{fig2}
\end{center}
\end{figure*}

\section{Theoretical Results--Band Structure and Superconductivity}
We now study the possibility of superconductivity in the tetragonal phase, which can be considered as natural due to the enhanced electronic hopping between the layers, which is likely to favor interlayer pairing superconductivity, as in La$_3$Ni$_2$O$_7$.
In Fig.~\ref{fig2}(b), we present the band structure of the six-orbital model of  La$_4$Ni$_3$O$_{10}$ at 40 GPa. Some of the key parameter values are listed in table~\ref{tab1}. For comparison, in Fig.~\ref{fig2}(a), we also show the band structure of the four-orbital model of La$_3$Ni$_2$O$_7$ obtained in our previous study~\cite{sakakibarala327}. While there are bonding (colored blue) and anti-bonding (green) $d_{3z^2-r^2}$ bands in La$_3$Ni$_2$O$_7$, in La$_4$Ni$_3$O$_{10}$, there is an additional non-bonding $d_{3z^2-r^2}$ band (yellow) between the bonding and anti-bonding bands~\cite{note-bond,Botana2022}. Interestingly, the Fermi level in La$_4$Ni$_3$O$_{10}$ is placed near the top of the bonding $d_{3z^2-r^2}$ band, while the bottom of the anti-bonding band is placed somewhat above the Fermi level, which is a situation similar to that in La$_3$Ni$_2$O$_7$. This similarity occurs due to a combination of two discrepancies between these materials, namely, (i) there is a non-bonding $d_{3z^2-r^2}$ band in La$_4$Ni$_3$O$_{10}$,  and (ii) the formal Ni valence is +2.67 in La$_4$Ni$_3$O$_{10}$ against +2.5 in La$_3$Ni$_2$O$_7$.

\begin{table}[!h]
\caption{The orbital level offset $\Delta E=E_{x^2-y^2}-E_{3z^2-r^2}$ between $d_{x^2-y^2}$  and $d_{3z^2-r^2}$ orbitals, the vertical interlayer hopping $t_{\perp}$ between the $d_{3z^2-r^2}$ orbitals, 
 the nearest-neighbor intralayer hoppings
   $t_{3z^2-r^2}$, $t_{x^2-y^2}$, and $t_{x^2-y^2{\text -}  3z^2-r^2}$ of  La$_4$Ni$_3$O$_{10}$  are displayed.
   Here, the onsite energy offset 
    between the inner- and outer-layer $d_{x^2-y^2}$ ($d_{3z^2-r^2}$) orbitals is 0.264 eV (0.416 eV).
   The first column indicates the layer difference of parameters.
   The parameters of  La$_3$Ni$_2$O$_7$ presented in Ref.~\cite{sakakibarala327} are also displayed in the lowest row for comparison.
\label{tab1}}
\begin{tabular}{c c c c c c c} \hline\hline
[eV]& \hspace{2pt}$\Delta E$ & \hspace{5pt}$t_{\perp}$ & \hspace{5pt}$t_{3z^2-r^2}$ & \hspace{2pt} $t_{x^2-y^2}$ & \hspace{5pt}$t_{x^2-y^2{\text -} 3z^2-r^2}$ \\\hline
 inner & \hspace{5pt}$0.102 $  \hspace{10pt}& \hspace{5pt}\multirow{2}{*}{$-0.715 $}  & \hspace{5pt}$-0.163 $ & \hspace{5pt}$-0.543 $ &\hspace{5pt} $-0.297$  \\
 outer & \hspace{5pt}$0.255 $  \hspace{10pt}& \hspace{5pt}                                         & \hspace{5pt}$-0.148$ & \hspace{5pt}$-0.539 $ &\hspace{5pt} $-0.285 $ \\
\hline
 La$_3$Ni$_2$O$_7$ & \hspace{5pt}$0.372$  \hspace{10pt}& \hspace{5pt}$-0.664$ & \hspace{5pt}$-0.117$ & \hspace{5pt}$-0.491$ &\hspace{5pt} $-0.242$ \\
\hline\hline
\end{tabular}
\end{table}

Let us now see how this similarity and discrepancies of the band structure between the two materials is reflected in superconductivity. In Fig.~\ref{fig2}(c), the eigenvalue of the Eliashberg equation $\lambda$ of the six-orbital model of La$_4$Ni$_3$O$_{10}$ is plotted against the band filling  (defined as the number of electrons per unit cell per spin), which is measured from that of the stoichiometric composition (two electrons). For comparison, we plot $\lambda$ for the four-orbital model of La$_3$Ni$_2$O$_7$~\cite{sakakibarala327}. The eigenvalue is found to be smaller for La$_4$Ni$_3$O$_{10}$ than for La$_3$Ni$_2$O$_7$, but a notable difference regarding the band filling dependence is that while $\lambda$ monotonically decreases for La$_3$Ni$_2$O$_7$ when the band filling is decreased, namely, when it moves away from half filling, for La$_4$Ni$_3$O$_{10}$, $\lambda$ is (locally) maximized near the stoichiometric composition.  Consequently, its value (0.433)  for La$_4$Ni$_3$O$_{10}$ at stoichiometry is not small, in the sense that it is still comparable to those of some of the cuprates (with relatively low $T_c$) obtained by the same method~\cite{sakakibara1,sakakibara2,sakakibara3}. The relatively small reduction of $\lambda$ compared to that of La$_3$Ni$_2$O$_7$ is even more interesting considering the fact that the band filling of the $d_{3z^2-r^2}$ orbitals in La$_4$Ni$_3$O$_{10}$ is only roughly 1/3 (per Ni atom per spin), compared to roughly 1/2 (half filling) in  La$_3$Ni$_2$O$_7$, so that the electron correlation effects are expected to be significantly smaller in the former.

\begin{figure}
\begin{center}
\includegraphics[width=7cm]{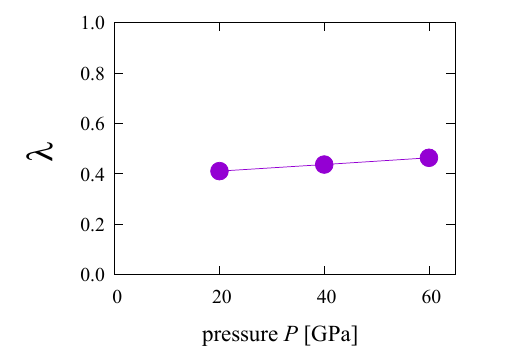}
\caption{The eigenvalue of the Eliashberg equation for La$_4$Ni$_3$O$_{10}$ plotted against the external pressure. \label{figp}}
\end{center}
\end{figure}

In order to understand the origin of the band filling dependence of $\lambda$, we  plot  in Fig.~\ref{fig2}(e) the superconducting gap function of the model of La$_4$Ni$_3$O$_{10}$ for various band fillings, together with that of La$_3$Ni$_2$O$_7$ (at stoichiometry) obtained in Ref.~\cite{sakakibarala327} (Fig.~\ref{fig2}(d)). At stoichiometry in La$_4$Ni$_3$O$_{10}$, the sign of the gap is reversed  not only between bonding and anti-bonding $d_{3z^2-r^2}$ bands (as in La$_3$Ni$_2$O$_7$), but also between bonding and non-bonding bands. When the band filling is decreased from stoichiometry (when holes are doped), the superconducting gap of the anti-bonding band becomes small, while when electrons are doped, the gap of the non-bonding band becomes small. Also, all the bands are fully gapped at stoichiometry, while the small gaps are nodal when doped. From these results, we may conclude that the (locally) maximized and relatively large $\lambda$ obtained for La$_4$Ni$_3$O$_{10}$ around stoichiometry, despite $d_{3z^2-r^2}$ orbitals being away from half filling,  is because all three $d_{3z^2-r^2}$ bands contribute to superconductivity. This in turn can be attributed to the relation between the Fermi level and the band edges;  around stoichiometry, the edge of all three bands touches or lies close to the Fermi level, namely all the bands are (nearly) ``incipient'', which pushes up the spin fluctuations to finite energies, thereby making them more effective as pairing glue~\cite{Nakata,KitaminePressure}. When holes are doped, the Fermi level moves away from the anti-bonding band bottom, whereas for electron doping, the Fermi level firmly intersects the non-bonding band. These situations are schematically presented in Fig.~\ref{fig2}(f).  It is also worth noting that the role played by the non-bonding band is quite different from that in the three-leg Hubbard ladder, where the non-bonding band is irrelevant for superconductivity~\cite{Kimura}. Although the origin of this discrepancy is not clear at present, there are differences in that $t_{\perp}$ is much larger than the in-plane hoppings and also the band filling is far away from half filling in the present model, besides the obvious differences in the dimensionality and the presence of the inter-orbital hybridization.
In the above, we have mainly focused on the gap function of the $d_{3z^2-r^2}$ orbitals, but as for the $d_{x^2-y^2}$  orbitals, the gap function takes relatively large values around the N point in all cases presented here, where the hybridization between the $d_{x^2-y^2}$ and $d_{3z^2-r^2}$ orbitals is strong. This tendency is similar to that found in our calculation for La$_3$Ni$_2$O$_7$ \cite{sakakibarala327}.

Finally, we study the pressure dependence of the superconductivity. We perform similar FLEX calculations also at 20 and 60 GPa, and obtain the eigenvalue $\lambda$. The eigenvalue is plotted against pressure in Fig.~\ref{figp}, which suggests that $T_c$ is expected to increase upon increasing the pressure.

\section{Experimental results--La$_3$Ni$_2$O$_{7.01}$}

\begin{figure*}
\begin{center}
\includegraphics[width=18 cm]{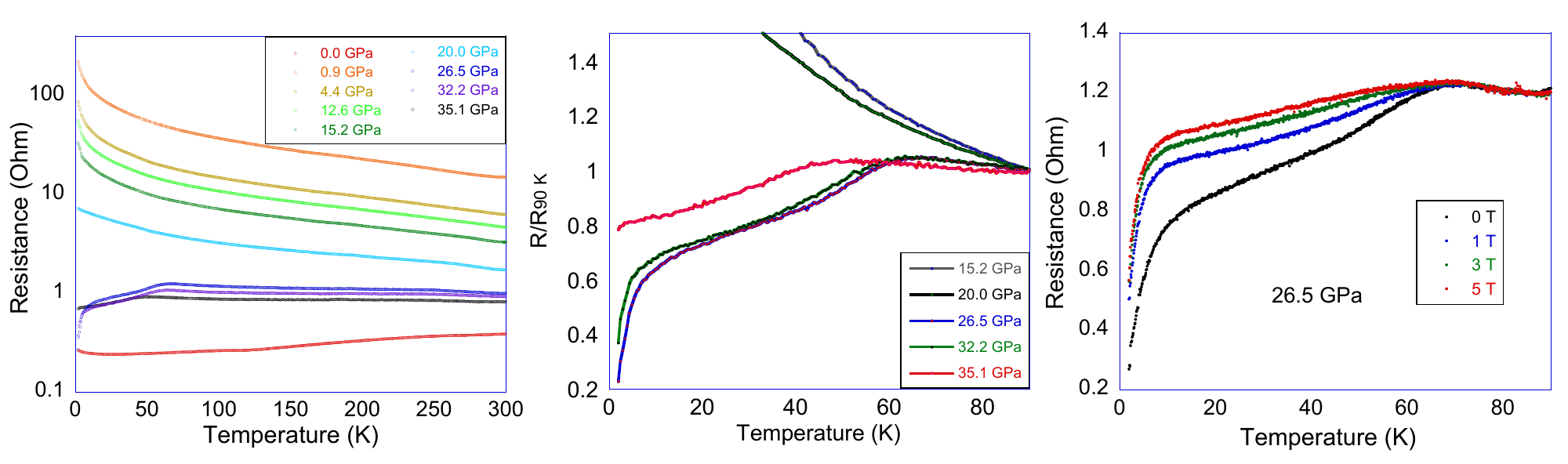}
\caption{Temperature dependence of electrical resistance of La$_3$Ni$_2$O$_{7.01}$ (a) under several pressures from 0.0 GPa to 35.1 GPa, (b) normalized at 90 K under 15.2 - 35.1 GPa, (c) under magnetic fields from 0 to 5 T at 26.5 GPa.}
\label{fig:3}
\end{center}
\end{figure*}

\begin{figure*}
\begin{center}
\includegraphics[width=18 cm]{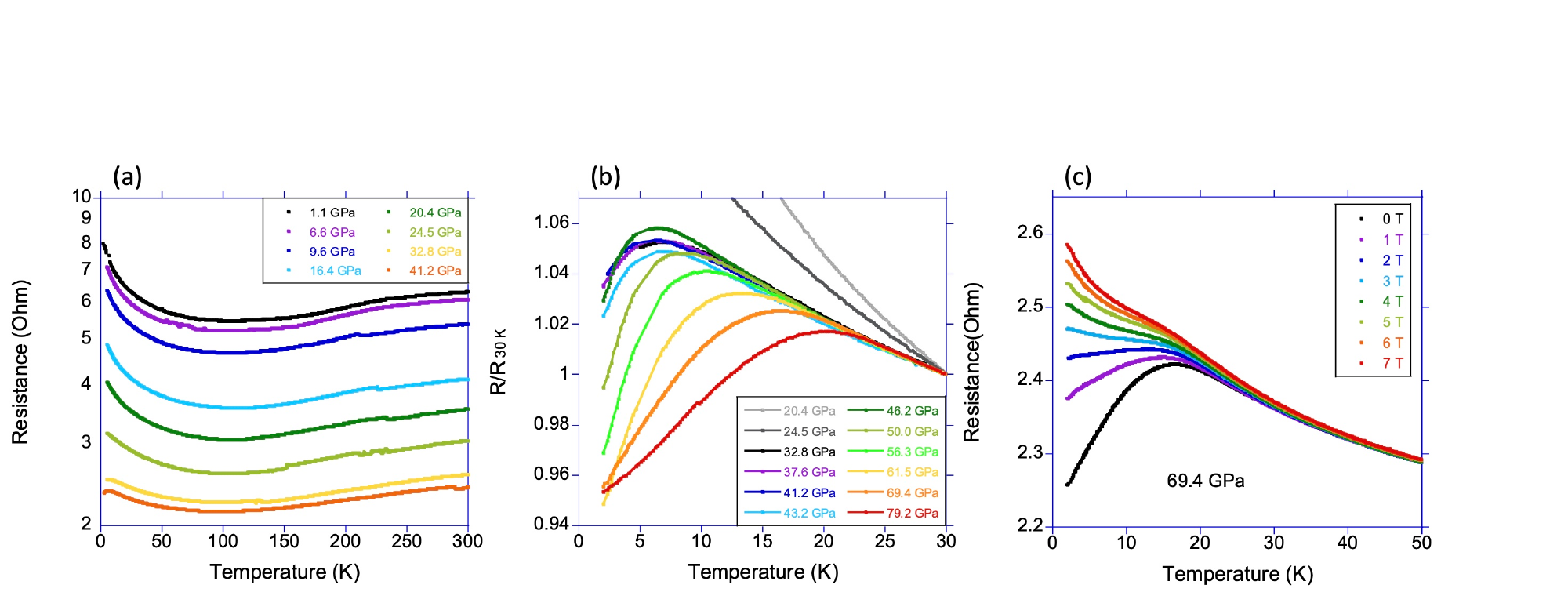}
\caption{Temperature dependence of the electrical resistance of La$_4$Ni$_3$O$_{9.99}$  (a) under several pressures from 1.1 to 41.2 GPa, (b) normalized at 30 K, under several pressures from 32.8 to 79.2 GPa, (c) under magnetic fields from 0 to 7 T, at 69.4 GPa }
\label{fig:4}
\end{center}
\end{figure*}

We now turn to the experimental results. We first focus on La$_3$Ni$_2$O$_{7.01}$. Fig.~\ref{fig:3} displays  the temperature dependence of the electrical resistance under various pressures and magnetic fields. 
Semiconducting behavior is observed below 20.0 GPa, as shown in Fig.~\ref{fig:3}(a), although it is likely caused by extrinsic factors such as grain boundary scattering because La$_3$Ni$_2$O$_{7-\delta}$ ($\delta \sim 0$) has been reported to exhibit metallic properties under ambient pressure~\cite{327resist}.
 At 26.5 GPa, the superconducting transition with a onset $T_c$ around 67.0  K is clearly recognized, which is a behavior similar to that of La$_3$Ni$_2$O$_7$ single crystals in the previous studies~\cite{MWang,2307.09865,2307.14819}, although the amount of the resistance drop is not as significant as in those studies. $T_c$ appears to decrease with increasing pressure as shown in Fig.~\ref{fig:3}(b). The superconductivity survives even at magnetic fields of 5 T although the resistance drop below  $T_c$ diminishes significantly with increasing field, as shown in Fig.~\ref{fig:3}(c). $T_c$ exhibits small field dependence, i.e., 67.0  K (0 T) to 65.3 K (5 T), which is also consistent with the previous studies ~\cite{MWang,2307.09865,2307.14819}. 

\section{Experimental results--La$_4$Ni$_3$O$_{9.99}$}
We now present experimental results for La$_4$Ni$_3$O$_{9.99}$. Fig.~\ref{fig:4} shows the temperature dependence of the electrical resistance under various pressures and magnetic fields. La$_4$Ni$_3$O$_{9.99}$ displays metallic behavior across all measured pressures, with a slight upturn observed at temperatures below approximately 100 K, as shown in Fig.~\ref{fig:4}(a). The origin of the upturn is uncertain at present. Intriguingly, the drop in resistance suddenly appears below 5 K at 32.8 GPa. Upon increasing the pressure beyond  46.2 GPa, the temperature at which the resistance is maximized increases, and also the decrease in the resistance below it becomes more pronounced, as seen in Fig.~\ref{fig:4}(b). Here, it should be reminded that according to our theoretical calculation, the tetragonal structure is stabilized beyond pressures around 20 GPa in La$_4$Ni$_3$O$_{10}$, for which superconductivity with a $T_c$ lower than that of La$_3$Ni$_2$O$_7$ is expected to take place around stoichiometric composition. 
Since the pressures where the reduction in the resistance is observed are well in the tetragonal phase regime, this reduction can most likely be attributed to pressure induced superconductivity of La$_4$Ni$_3$O$_{9.99}$, 
although the magnitude of the resistance drop is not as significant as in single crystal La$_3$Ni$_2$O$_7$~\cite{MWang,2307.09865,2307.14819}, similarly to our data for polycrystalline La$_3$Ni$_2$O$_{7.01}$. In fact, the magnetic field dependence of the resistance is quite similar to  that of La$_3$Ni$_2$O$_{7.01}$ as shown in \ref{fig:4}(c), namely, the peak position of the resistance is nearly unaffected, while the resistance drop becomes less significant with increasing field. On the other hand, as seen in Fig.~\ref{fig:4}(b), the temperature at which the resistance deviates from a linear behavior, considered as the onset $T_c$,  monotonically increases up to 23 K upon increasing the  pressure up to 79.2 GPa, which is a  pressure dependence opposite to what is observed in La$_3$Ni$_2$O$_{7.01}$  and hence is characteristic of La$_4$Ni$_3$O$_{9.99}$.
This tendency may also be considered as consistent with the theoretical results, where the eigenvalue of the Eliashberg equation increases with pressure (Fig.~\ref{figp}).

\section{Conclusion}
In the present study,  we have investigated the possibility of superconductivity in La$_4$Ni$_3$O$_{10}$ under pressure both theoretically and experimentally. Through DFT calculations, we have found that a structural phase transition from monoclinic to tetragonal takes place around 20 GPa. Using the crystal structure obtained at 40 GPa, we have theoretically investigated the possibility of superconductivity, where a combination of FLEX and linearized Eliashberg equation is applied to a six-orbital model constructed from first principles band calculation. An interesting feature found here is that the eigenvalue of the Eliashberg equation is (locally) maximized around stoichiometry, reflecting the fact that all of the three $d_{3z^2-r^2}$ bands contribute to superconductivity around this band filling. 
This is due to the fact that all the bands are (nearly) incipient, which pushes up the spin fluctuations to finite energies and make them more effective as pairing glue.
Hence our calculation results suggest that La$_4$Ni$_3$O$_{10}$, around stoichiometric composition, may become superconducting under high pressure with $T_c$ comparable to some cuprates, although it is not as high as La$_3$Ni$_2$O$_7$. In our analysis, we have assumed that superconductivity, if any, occurs in the tetragonal structure phase. We speculate that other orders such as the charge density wave is more favored for lower symmetries, but such a competition between superconductivity and other orders is beyond the scope of the present study, and serves as an interesting future issue.

We have also examined our theoretical analysis by performing experimental studies using our polycrystalline samples of La$_3$Ni$_2$O$_{7.01}$ and La$_4$Ni$_3$O$_{9.99}$. The superconducting transition of La$_3$Ni$_2$O$_{7.01}$ has been confirmed by a drop in the electrical resistance, as well as the magnetic field dependence of the resistance. We have observed a maximum onset $T_c$ of 67.0 K at a pressure of 26.5 GPa. Similar temperature and magnetic field dependencies of the resistance have been observed also for La$_4$Ni$_3$O$_{9.99}$ under high pressure, where a drop in the resistance is observed at lower temperatures compared to La$_3$Ni$_2$O$_{7.01}$. Given the theoretical expectation, the reduction in the resistance can most likely be attributed to the occurrence of superconductivity in La$_4$Ni$_3$O$_{9.99}$, although we believe that further studies are necessary for a complete confirmation. The onset $T_c$, defined as the temperature at which the resistance deviates from a linear behavior, increases with pressure up to 23 K at 79.2 GPa, which is the opposite of what is observed for La$_3$Ni$_2$O$_{7.01}$. Understanding the origin of this difference between the two materials also serves as an interesting future study.

\ \\
During the finalization process of the present study, we came to notice a recent experimental study on polycrystalline samples~\cite{2309.01651}. The temperature and magnetic field dependencies of the resistance in La$_3$Ni$_2$O$_7$ observed there are quite similar to our results on La$_3$Ni$_2$O$_{7.01}$, while superconductivity is not observed in La$_4$Ni$_3$O$_{10}$ under  the pressure of up to 50 GPa.
Also, after the initial submission and before the resubmission of our manuscript, there appeared three studies that suggest occurrence of superconductivity in La$_4$Ni$_3$O$_{10}$ under pressure. In Ref. \cite{2311.05453}, results similar to ours are obtained for polycrystalline samples. In Refs. \cite{2311.07423,2311.07353}, single crystals are studied, and more clear drops of resistance are obtained at around 20-25 K at high pressures.
\ \\

\begin{acknowledgments}
We thank Kenji Kawashima for fruitful discussions.
We are supported by JSPS KAKENHI Grant No. JP22K03512 (H. Sakakibara), JP22K04907 (K. K.), 
JP20H05644 (Y. T.), and JSPS Bilateral Program JPJSBP120214602 (Y. T.).
The computing resource is supported by 
the supercomputer system HOKUSAI in RIKEN,
the supercomputer system (system-B) in the Institute for Solid State Physics, the University of Tokyo, 
and the supercomputer of Academic Center for Computing and Media Studies (ACCMS), Kyoto University.

H.S.(H. Sakakibara), M.O., H.N., Y.U., and H.S.(H. Sakurai) contributed equally to this work.
\end{acknowledgments}

\bibliography{la4310}

\begin{thebibliography}{104}
\bibitem [{not({\natexlab{a}})}]{note-int}%
  \BibitemOpen
  \href@noop {} {}  \bibinfo {note} {For La$_3$Ni$_2$O$_7$,
  we also performed FLEX calculation adopting a more realistic set of
  interaction values evaluated by constrained-RPA \cite{WernercRPA}, and
  obtained similar results.}\BibitemShut {Stop}%
\end{thebibliography}

\end{document}